\documentclass[a4paper,11pt]{article}
\usepackage{jcappub} 
\usepackage{lineno}
\usepackage[compat=1.1.0]{tikz-feynman}

\title{\boldmath 
Less structure on $8$ Mpc scales from decaying sterile neutrino dark matter
}

\author[a]{Mar\'{i}a Dias Astros}
\affiliation[a]{Institute of Physics, University of Freiburg, Hermann-Herderstr. 3, 79102 Freiburg, Germany}

\author[b,c]{, Luk\'{a}\v{s} Gr\'{a}f}
\affiliation[b]{Institute of Particle and Nuclear Physics, Faculty of Mathematics and Physics, Charles University in Prague, V Holešovičkách 2, 180 00 Praha 8, Czech Republic}
\affiliation[c]{Theory Group, Nikhef, Science Park 105, 1098 XG, Amsterdam, The Netherlands}

\author[a]{and Stefan Vogl}

\emailAdd{maria.dias@physik.uni-freiburg.de}
\emailAdd{lukas.graf@matfyz.cuni.cz}
\emailAdd{stefan.vogl@physik.uni-freiburg.de}

\abstract{
Late decays of dark matter to a lighter, warm dark matter component are a known way to reduce the amplitude of the matter power spectrum on scales of $8$ Mpc. However,  only very few particle physics models have been put forward that exhibit the required properties and allow to relate them to other observables. In this work, we investigate a model based on two interacting sterile neutrinos and a scalar singlet. The  heavier of the neutrinos is produced in the early Universe by the interplay of oscillations and the new interactions in the dark sector and constitutes the dominant component of dark matter. If the Yukawa matrix that describes the interactions of the steriles with the scalar is not diagonal, the heavier state can decay to three light sterile neutrinos. In contrast to the usual scenario, this leads to an all massive final state without radiation-like particles. We identify the part of the parameter space where these decays can lead to a reduction of S$_8$ at a level that matches observations. We then confront this region with the requirements of reproducing the observed relic density, as well as existing constraints from X-ray searches and Lyman-$\alpha$ forest data.
}

\begin{document}
\maketitle
\flushbottom

\section{Introduction}

The cold, collisionless, and stable dark matter (DM) paradigm has long served as the cornerstone of modern cosmology, successfully reproducing key observations such as the anisotropies of the Cosmic Microwave Background (CMB) and the distribution of large-scale structure in the Universe. Nonetheless, neither theoretical considerations nor observational evidence rule out departures from this minimal picture. In fact, numerous extensions of the Standard Model (SM) predict dark sectors containing additional particles and interactions that can naturally give rise to unstable or warm dark matter components.

While current cosmological and astrophysical data impose stringent limits on DM decays into SM particles, see e.g.~\cite{Poulin:2016anj,Yuksel:2007dr,Zhang:2007zzh,Slatyer:2016qyl,Cohen:2016uyg} or \cite{Cirelli:2024ssz} for a recent review with a comprehensive overview of current searches for DM decay, the situation is markedly different for decays proceeding entirely within the dark sector~\cite{Scherrer:1984fd,Poulin:2016nat,Nygaard:2020sow,Alvi:2022aam}. Such processes remain comparatively unconstrained, yet they can leave observable imprints on the evolution of cosmological perturbations and the matter power spectrum. 
Decays that occur at late times, i.e long after matter–radiation equality, can inject kinetic energy into the dark matter population, altering the growth of structure and potentially alleviating tensions between low- and high-redshift probes of $\Lambda$CDM cosmology. While the success of the these models in addressing the Hubble tension is limited~\cite{Haridasu:2020xaa,Clark:2020miy,Schoneberg:2021qvd},  they are quite successful in reducing the late-time value of $S_8$  \cite{FrancoAbellan:2020xnr,FrancoAbellan:2021sxk,Bucko:2024izb,Davari:2022uwd,Fuss:2022zyt} provided that some of the decay products are produced non-relativistically. 
Despite the considerable cosmological interest in such solutions, particle physics models that realize them have been investigated sparingly \cite{Choi:2021uhy,Fuss:2024dam,Cheek:2024fyc,Fuss:2025xwe}. We aim to address this gap in this work.

A particularly appealing dark matter candidate are sterile neutrinos. These singlet fermions can interact with the SM via a dimension four interaction with the active neutrinos and the Higgs. In the broken phase, this leads to mixing between the active and the sterile neutrinos that allows for the production of the sterile neutrinos  by an interplay of collisions and oscillations as first suggested by Dodelson and Widrow \cite{Dodelson:1993je}. 
This production mechanism is very appealing due to its simplicity and minimal field content, but, unfortunately, the preferred parameter space for dark matter is excluded by other searches. 
Nevertheless, there are various extensions of this idea that allow to preserve some of the attractive features, e.g. resonant enhancement of the oscillation probability by a large lepton asymmetry \cite{Shi:1998km,Abazajian:2001,Kishimoto:2006zk}, novel interactions of active neutrinos \cite{DeGouvea:2019wpf,Kelly:2020pcy,Benso:2021hhh}, or thermalization of the sterile neutrinos \cite{Hansen:2017rxr}. An intriguing alternatives arise with the sterile neutrino production boosted by self-interactions~\cite{Johns:2019cwc,Bringmann:2022aim,Astros:2023xhe}.
As the cosmological solutions we are after require two long-lived particle species in the dark sector and interaction between them, we focus on an extended version of models that realize sterile neutrino self-interactions as these already include
a mediator that can enable the interactions between the dark matter particles. Concretely, 
we focus on a model in which a heavier sterile neutrino dark matter state decays into a lighter sterile neutrino at late cosmological times. The resulting population of mildly boosted daughter particles effectively behaves as a warm or mixed dark matter component, leading to a suppression of power on small and intermediate scales. We construct a minimal framework where the sterile neutrino decay is mediated by a massive scalar, for an EFT realization of this idea see e.g. \cite{Fuller:2024noz}. The same interactions responsible for the decay are also crucial for the generation of sterile neutrinos in the early Universe, where they enhance the usual production via scattering-induced decoherence. Thus, the studied scenario opens the possibility of a unified description of dark matter genesis and its late-time phenomenology.

The paper is structured as follows. In Sec.~\ref{sec:decay_overview} we briefly review the status of dark matter decaying to warm dark matter and its implications for structure formation. Next, in Sec.~\ref{sec:DM_production}, we introduce our model for interacting sterile neutrinos and discuss how the right relic density can be produced by an interplay between mixing with active neutrinos and self-interactions in the sterile neutrino sector. We close this section with a study of the decay of the sterile neutrino dark matter candidate to a lighter sterile neutrino species which takes the role of the warm dark matter component in our model. In Sec.~\ref{sec:our_decays}, we combine our results for the lifetime and the energy release with the relic density computation to identify the part of the parameter space that leads to a reduction of the matter power spectrum at a level consistent with observations. Finally, we summarize our results in Sec.~\ref{sec:summary}.

\section{Decays to warm dark matter and cosmic structure}
\label{sec:decay_overview}

The effect of cold dark matter decaying to dark radiation on cosmological observables has been studied intensely since the days of WMAP, see e.g. \cite{Ichiki:2004vi,DeLopeAmigo:2009dc,Audren:2014bca}. As the data has improved, this has allowed to place increasingly stringent limits on the lifetime of dark matter and the fraction of dark matter that is allowed to undergo such decays. Current analysis are able to exclude lifetimes $\tau$ shorter than $\sim 250$ Gyr \cite{Nygaard:2020sow,Alvi:2022aam} if the decaying component dominates the relic density. 
When all the decay products are ultra-relativistic, these decays can affect the cosmological perturbations on quite large scales, and the limits are driven by Planck observations of the CMB. If one or more of the particles produced in the decay have a non-negligible mass, the picture changes. In this case, the non-relativistic velocity of the massive daughter particle introduces a new length scale in the problem. In a way that is reminiscent of the free-streaming length of warm dark matter models: a characteristic length scale for the suppression of structure arises. This makes large-scale structure surveys, which target scales smaller than the CMB, more sensitive. 
In addition, for lifetimes of the order of the age of the Universe or longer  the impact on the CMB is suppressed, because most decays occur after photon decoupling. This has led to interest in decays to massive particles since it may alleviate the longstanding discrepancies in the amplitude of the matter power spectrum on scales of $\sim 8$ Mpc, conventionally measured by the quantity S$_8$. 
While Planck prefers a value of S$_8= 0.832 \pm 0.013$ \cite{Planck:2018vyg} other determinations tend to prefer a lower value, see e.g.~\cite{Abdalla:2022yfr,CosmoVerseNetwork:2025alb} for two recent review papers of tensions in cosmology that include an extensive discussion of the various S$_8$ determinations. For example, DES measures S$_8=0.776\pm0.017$~\cite{DES:2021vln} or S$_8=0.772 \pm 0.018$~\cite{DES:2021wwk} depending on the methodology, while HSC prefers a similar value of S$_8=0.776_{-0.033}^{+0.032}$ \cite{Dalal:2023olq} albeit with a larger error. 
Note, however, that some studies such as the KiDS Legacy analysis found a higher value of S$_8=0.814^{+0.011}_{-0.012}$ 
\cite{Stolzner:2025htz} which is broadly consistent with Planck.
Even though the evidence for an S$_8$ tension is less clear than for the Hubble tension and only amounts to $\sim 2 \sigma$, it is still interesting to investigate what properties are required to resolve it and how these can be realized in a particle physics model.

A minimal realization of the decaying dark matter solution for both the Hubble and the S$_8$ tension, which has been studied in the literature in considerable detail \cite{Haridasu:2020xaa,FrancoAbellan:2020xnr,Clark:2020miy,Schoneberg:2021qvd,FrancoAbellan:2021sxk,Bucko:2024izb,Davari:2022uwd,Fuss:2022zyt}, consists of a dark matter particle $X$ decaying to a slightly lighter companion $X_{WD}$, that acts as warm dark matter, and a light particle $\gamma_{D}$ which can be treated as dark radiation. The decay 
\begin{align*}
    X \rightarrow X_{WD} + \gamma_{D}
\end{align*}
 is characterized completely by the lifetime $\tau$ and the released energy. The energy release can be quantified by parameter $\epsilon=p/M=\frac{1}{2}(1-\frac{m^2}{M^2})$, which gives the momentum of $X_{WD}$ as a fraction of the mass of the decaying particle. The velocity of the massive decay daughter at production is $\beta_{WD}=\sqrt{\epsilon^2/(1-\epsilon^2)}\approx \epsilon$ for $\epsilon\ll 1$. Similar to standard warm dark matter, this velocity controls the distance that these particles will travel after production and, thus, $\epsilon$ sets the scale at which a suppression of structure sets in.
 A value of S$_8\approx0.77$, as preferred by DES and HSC, is found for values $\epsilon= 0.1 \-- 0.01$ and $\tau= 4 \-- 10 \times 10^{18}$ s \cite{Fuss:2024dam}.

Relatively few concrete particle physics models that allow to track the dynamics of such a  decaying dark matter candidate in the early Universe and study its implications for other observables have been put forward. 
In the set-up described above, three particles beyond the SM are required: the decaying DM particle, the warm dark matter daughter particle, and a light degree of freedom that acts as radiation.
A construction that realizes this is, for example, discussed in \cite{Cheek:2024fyc}.
Note, however, that one can keep the relevant features with fewer new particles. One possibility is to identify the radiation with SM neutrinos \cite{Choi:2021uhy} or consider a three-body decay where a pair of neutrinos \cite{Fuss:2024dam,Fuss:2025xwe} takes the role of the dark radiation\footnote{Indirect detection limits on the production of other SM particles are typically too strong to allow for a decay with the required lifetime, see e.g.~\cite{Cirelli:2024ssz}.}.  In any case, the minimal requirements are two long-lived states with masses chosen such that the momenta of the daughter particle(s) are in the non-relativistic regime. Alternatively, one can consider a situation where only warm dark matter is produced in the decay without any accompanying radiation. To the best of our knowledge, this has not been considered in the literature before.
In this work, we will study such a scenario and investigate a model based on two interacting sterile neutrinos that allows for a suppression of structure on the scales relevant for S$_8$ via a decay to three massive particles.

\section{Production and decay of sterile neutrino dark matter}
\label{sec:DM_production}
In this study, we consider a framework with two sterile neutrinos that interact through self-interactions mediated by a singlet scalar field, $\phi$, via a Yukawa-type coupling    \begin{equation}
-\mathcal{L}_\text{int} = y_{ij} ~\bar{N}_i N_j \phi.
\label{eq: model lagrangian}
\end{equation} 
In addition, both sterile neutrinos can mix with the active neutrinos.
In our setup, we assume that the heavier state, $N_1$, mixes more strongly with the active neutrinos while the mixing of the lighter state is taken to be negligible. For simplicity, we restrict ourselves to mixing with the electron neutrino only, characterized by a single mixing angle $\theta$.

The coupling structure in Eq.\,\eqref{eq: model lagrangian} allows a number of processes that can be relevant for the cosmological evolution of the sterile neutrino populations.
First, the mixing with the active neutrinos allows the production of the $N_1$ state via the Dodelson-Widrow mechanism \cite{Dodelson:1993je}. Once a sufficient seed population of $N_1$ has been produced in this way, the interaction with the scalar leads to self-interactions of $N_1$. These participate in the collision induced decoherence of the active neutrinos and $N_1$ and thus contribute to the source term for the production of $N_1$ from oscillations \cite{Astros:2023xhe,Bringmann:2022aim}. 
Heuristically, this can be understood as a $\nu N_1 \rightarrow N_1 N_1$ scattering which is suppressed compared to the self-scattering $N_1 N_1 \rightarrow N_1 N_1$ by the insertion of the small mixing angle $\theta.$  
Since this production mode depends on the density of $N_1$ particles, it self-accelerates and can boost the dark matter relic density significantly compared to the expectation of a pure Dodelson-Widrow production. This is similar to the boost of the production found in more general freeze-in  models with semi-annihilations \cite{Bringmann:2021tjr,Hryczuk:2021qtz}.
We are interested in the region of parameter space where $N_1$ plays the role of dark matter and sketch the production mechanism in more detail in the next subsection. 
Second, the interactions of $N_2$ with the scalar allow for a scattering induced transfer of the $N_1$ population to the $N_2$ state as discussed in an effective operator limit in \cite{Fuller:2024noz}. 
As we are interested in late decays of $N_1$, we need to ensure that this process is inefficient such that a significant $N_1$ population survives until the present. We investigate the conditions for this to hold in Sec.~\ref{subsec:conversion}. Last but not least, the couplings $y_{12}$ and $y_{22}$ enable the decay $N_1\rightarrow 3 N_2$.
This decay produces non-relativistic decay products that can take the role of the warm dark matter component in the conventional two-body decay if the kinetic energy released in the decay, $E_{kin}=m_1 - 3\, m_2$, is small compared to $m_1$. We will discuss the decay width and the restrictions on the masses required by getting a suppression of S$_8$ on the right scales in Sec.~\ref{sec:width}. Note that the additional processes enabled by the $y_{22}$ and $y_{12}$ couplings were not considered in \cite{Astros:2023xhe,Bringmann:2022aim}.

\subsection{Production}
\label{sec: boosted production}

The production mechanism of $N_1$ in the early Universe is similar to the one described in \cite{Astros:2023xhe}, which we briefly review now. 
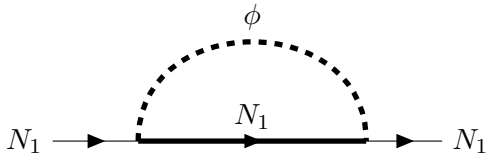
\begin{figure}[t]
\centering
\begin{tikzpicture}
\begin{feynman}
\vertex (a1) {\(N_1\)};
\vertex[right=1.5cm of a1] (a2);
\vertex[right=3cm of a2] (a3);
\vertex[right=1.0cm of a3] (a4) {\(N_1\)};

\diagram* {
{[edges=fermion]
(a1) -- [] (a2) -- [edge label=\(N_1\), line width = 0.7mm] (a3) --  (a4),
},
(a2) -- [scalar, ,out=90, in=90, looseness=1.5, edge label=\(\phi\), line width = 0.7mm] (a3),

};
\end{feynman}
\end{tikzpicture}
\caption{Lowest-order contributions to the sterile neutrinos' self-energy. Thick lines represent thermal propagators.}
\label{fig: lowest order correction to the neutrinos' self-energy}
\end{figure}

%
Assuming that there are no sterile neutrinos prior to their production through the mechanism we describe in the following, the evolution of the phase-space density $f_{N_1}$ is given by the Boltzmann equation  
\begin{equation}
\frac{\partial f_{N_1}(t,p)}{\partial t} - H p \frac{\partial f_{N_1}(t,p)}{\partial p} = \frac{\Gamma_t}{4}\langle P_m(N_1 \leftrightarrow \nu_e)\rangle f _e(p,t) + \mathcal{C}_{N_1},
\label{eq: Boltzmann eq}
\end{equation} 
where $H$ is the Hubble rate, $f_e$ is a Fermi-Dirac distribution function for the active neutrinos and $\Gamma_t = \Gamma_a + \Gamma_{N_1}$ is the total (active plus sterile) rate.  Pauli blocking has been neglected here since we are interested in a situation where the neutrino gas is dilute.  The transition probability is defined as
\begin{equation}
\left< P_m(N_1 \leftrightarrow \nu_e)\right> = \frac{\omega^2(p) \sin^2(2 \theta)}{\omega^2(p) \sin^2(2\theta) + D^2(p) + \left[\omega(p) \cos(2\theta) - V_{\text{eff}}\right]^2},
\end{equation}
with $\omega (p) \approx m_1 / 2p$ the vacuum oscillation frequency. In-medium effects are encapsulated in the quantum damping term, defined as $D(p) = \Gamma_t/2$ and the effective potential $V_\text{eff}$. The latter is induced by forward scattering of the neutrinos in the plasma acting as a thermal correction to the neutrino's self-energy that appears at one loop (see Fig.\,\ref{fig: lowest order correction to the neutrinos' self-energy}). This potential comprises a SM and a sterile contribution, i.e. $V_\text{eff} = V_a - V_{N_1}$. Finally, the term $\mathcal{C}_{N_1}$ denotes a sterile collision term that takes the form 
 \begin{equation}
    \mathcal{C}_{N_1} (p) = -\Gamma_{N1 N1 \leftrightarrow \phi \phi} \, f_{N_1}(p) \left( \frac{f_{N_1}^2 (p)}{f_{\text{eq}}^2(p)}-1 \right),
    \label{eq: number changing collision term}
\end{equation}
where $f_{\text{eq}}(p) = e^{-p/T_s}$ is the equilibrium distribution function and $\Gamma_{N_1 N_1 \leftrightarrow \phi \phi}$ denotes the rate for $\phi$ pair production. Notice that this process results in the production of four neutrinos $N_1$ following the decay of the scalars. Full expressions for the potentials and the rates, along with more details, can be found in \cite{Astros:2023xhe}.

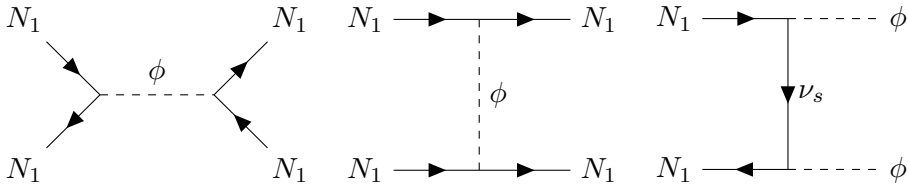
\begin{figure}[t]
\centering
\begin{tikzpicture}
\begin{feynman}
\vertex (a1) {\(N_1\)};
\vertex[below = 2cm of a1] (a2){\(N_1\)};
\vertex at ($(a1)!0.5!(a2) + (1, 0)$)[dot] (a3);
\vertex[right=1.5cm of a3] (a4);
\vertex[right = 3.5cm of a1] (a5){\(N_1\)};
\vertex[below = 2cm of a5] (a6){\(N_1\)};
\diagram* {
{[edges=fermion]
(a1) -- [] (a3) -- [] (a2),
},
(a3) -- [scalar, edge label=\(\phi\)] (a4),
(a4) -- [fermion] (a5), 
(a6)-- [fermion] (a4)
};
\end{feynman}
\end{tikzpicture}
\hspace*{0mm}
\begin{tikzpicture}
\begin{feynman}
\vertex (a1) {\(N_1\)};
\vertex[right= 1.5cm of a1](a3);
\vertex[right = 1.2cm of a3] (a2){\(N_1\)};
\vertex[below=2cm of a3] (a4);
\vertex[right = 1.2cm of a4] (a5){\(N_1\)};
\vertex[below = 2cm of a1] (a6){\(N_1\)};
\diagram* {
(a1) -- [fermion] (a3),
(a3) -- [fermion] (a2),
(a3) -- [scalar, edge label=\(\phi\)] (a4),
(a4) -- [fermion] (a5), 
(a6) -- [fermion] (a4)
};
\end{feynman}
\end{tikzpicture}
\hspace*{0mm}
\begin{tikzpicture}
\begin{feynman}
\vertex (a1) {\(N_1\)};
\vertex[right= 1.5cm of a1](a3);
\vertex[right = 1.2cm of a3] (a2){\(\phi\)};
\vertex[below=2cm of a3] (a4);
\vertex[right = 1.2cm of a4] (a5){\(\phi\)};
\vertex[below = 2cm of a1] (a6){\(N_1\)};
\diagram* {
(a1) -- [fermion] (a3),
(a3) -- [scalar] (a2),
(a3) -- [fermion, edge label=\(\nu_s\)] (a4),
(a4) -- [scalar] (a5), 
(a4)-- [fermion] (a6)
};
\end{feynman}
\end{tikzpicture}
\caption{Representative diagrams for the sterile neutrino processes relevant for DM production.}
\label{fig:relevant scattering processes for the production of DM.}
\end{figure}

Note that, compared to the minimal Dodelson-Widrow scenario \cite{Dodelson:1993je}, the introduction of sterile neutrino self-interactions modifies the system in two key ways: (i) through additional production and annihilation channels, exemplified by processes like $N_1 + N_1 \leftrightarrow N_1 + N_1$ and $N_1 + N_1 \leftrightarrow \phi + \phi$ (see Fig.\,\ref{fig:relevant scattering processes for the production of DM.}), and (ii) with a new thermal correction to the sterile neutrinos' dispersion relation arising from the forward scattering in the plasma (see Fig.\,\ref{fig: lowest order correction to the neutrinos' self-energy}).   

With these modifications, the production of sterile neutrino dark matter is determined by the interplay between the interaction rates, the effective potential, and the vacuum oscillation frequency. The relative importance of these contributions depends primarily on the mediator mass and the coupling strength $y_{11}$. In fact, one can identify different production regimes that are characterized by different phenomenology. In particular, for low mediator masses, an initially DW produced population of $N_1$'s is boosted by the presence of the new interactions and,  for sufficiently large couplings, an additional enhancement may occur as the system approaches thermalization, driven by number-changing processes. For heavier mediators, on the other hand, coherent processes might be important and resonant production can happen in a way analogous to the 
MSW effect \cite{Wolfenstein:1977ue, Mikheyev:1985zog}.  In both cases, the final dark matter abundance can be orders of magnitude higher than the seed populations, such that mixing angles $\sin \theta$ much smaller than those required by the Dodelson-Widrow mechanism can explain the observed relic density.
Concretely, for mediator masses in the range of a few keV to $\sim 3 \,\text{GeV}$, the observed dark matter relic abundance can be achieved in the $2 - 100\,\text{keV}$ mass window for mixing angles that are at least three orders of magnitude smaller than those required by the Dodelson–Widrow mechanism (see \cite{Astros:2023xhe} for more details). 

\subsection{Conversion}
\label{subsec:conversion}
The interaction given in Eq.\,\eqref{eq: model lagrangian} also opens up the possibility of two $N_1$ particles annihilating into two $N_2$ particles via an s-channel $\phi$ exchange. We assume that there is a mild hierarchy between the couplings such that this process does not impact the production of $N_1$ at high temperatures. However, even in this case the conversion might become efficient at low temperatures, i.e. when $T$ drops to about the mass of $N_1$. If the conversion is efficient, it can significantly deplete the population of $N_1$'s. 
This is problematic, since a non-negligible abundance of the heavier component must survive until late times in order to decay. 
Using the fact that the total yield $Y = Y_1 + Y_2$, with $Y_i = n_i/s$ where $n_i$ is the number density of species $i$ and $s$ is the entropy density, is conserved during this process, we can describe the evolution by a single Boltzmann equation for $Y_1$. It reads 
\begin{equation}
    \frac{\text{d} Y_1}{\text{d}x} = - \sqrt{\frac{\pi}{45}} g_\star^{1/2} M_\text{Pl}\frac{m_1}{x^2} \langle \sigma v \rangle \left[Y_1^2 - \left(\frac{Y_1^\text{eq}}{Y_2^\text{eq}}\right)^2 (Y-Y_1)^2\right],
    \label{eq: boltzmann equation for conversion}
\end{equation}
where $Y_{i}$ $ (Y_{i}^\text{eq})$ are the (equilibrium) yields, $M_\text{Pl}$ is the Planck mass and we have defined $x \equiv m_1/T_\gamma$ with $T_\gamma$ the SM temperature. Here, $g_\star^{1/2}$ is given by 
\begin{equation}
    g_\star^{1/2}(T_\gamma) = \frac{h_\text{eff}(T_\gamma)}{g_\text{eff}^{1/2}(T_\gamma)}\left(1+ \frac{T_\gamma}{3 h_\text{eff}}\frac{\text{d}h_\text{eff}}{\text{d}T_\gamma}\right),
\end{equation}
with $g_\text{eff}$ and $h_\text{eff}$ the effective number of degrees of freedom in energy and entropy, respectively. We use the tabulated values reported in \cite{Laine:2015kra,Laine_eos} to evaluate $g_\text{eff}$ and $h_\text{eff}$. Approximating the $N_1$ phase space density as a Maxwell-Boltzmann distribution, i.e. $f_1 \propto e^{-E_1/T_s}$, the thermally averaged cross section is  \cite{Gondolo:1990dk} 
\begin{equation}
    \langle \sigma v\rangle = \frac{1}{8m_1^4K_2^2(m_1/T_s)} \int_{4m_1^2}^\infty \sigma \left(s- 4m_1^2\right) \sqrt{s} K_1(\sqrt{s}/T_s) \, \text{d}s,
    \label{eq: thermally averaged cross section}
\end{equation}
where $K_{1,2}$ are the modified Bessel functions of the second kind, $\sigma$ is the cross section, and $s$ is the center-of-mass energy.
Note that in Eq.\,\eqref{eq: thermally averaged cross section} we allow the sterile sector to have its own temperature $T_s$ which, in the most general case, differs from that of the SM plasma. Furthermore, the right-hand side of Eq.\,\eqref{eq: boltzmann equation for conversion} drives the system towards chemical equilibrium between $N_1$ and $N_2$ and it vanishes once $Y_1 = \frac{Y_1^\text{eq}}{Y_2^\text{eq}} Y_2$. As $Y_1^\text{eq} / Y_2^\text{eq}\ll 1$ for $T< m_1$, this can significantly deplete the population of $N_1$. 

\begin{figure}
    \centering
    \includegraphics[width=0.7\linewidth]{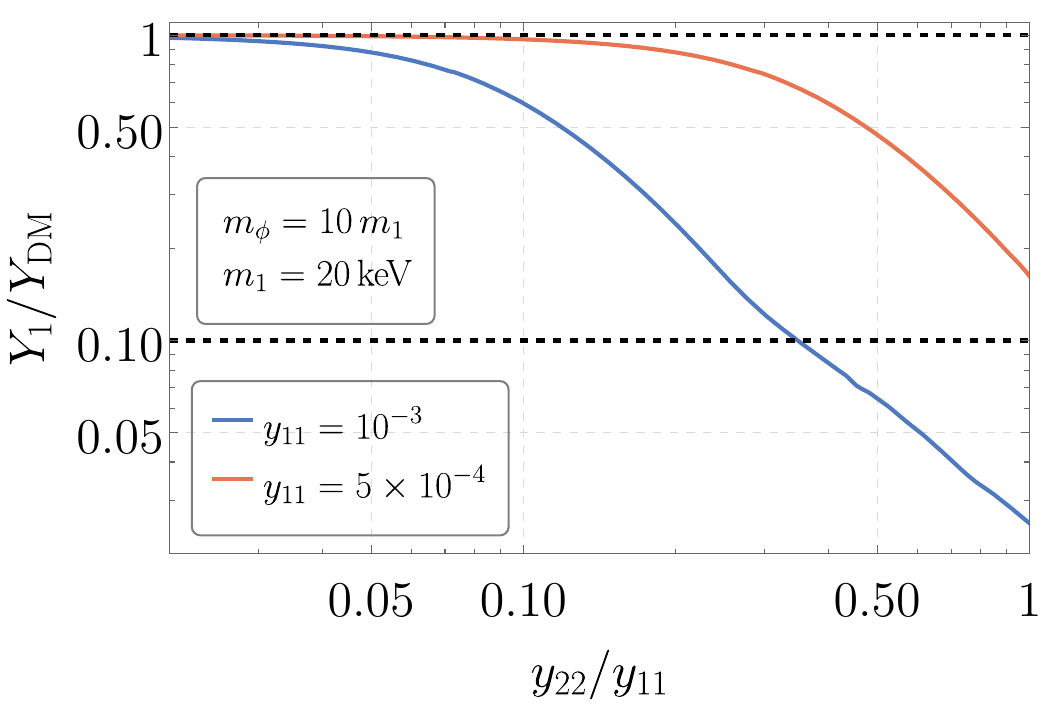}
    \caption{Evolution of the fraction of $N_1$ particles as a function of the ratio $y_{22}/y_{11}$ for an illustrative mass ratio $m_\phi/m_1 = 10$. The blue solid line corresponds to $y_{11} = 10^{-3}$, while the orange solid line corresponds to $y_{11} = 5 \times 10^{-4}$. The dashed black lines indicate the fractions corresponding to $N_1$ accounting for the entire dark-matter relic abundance and for 10\% of it.}
    \label{fig:conversion}
\end{figure}

In particular, we are interested in scenarios where at least $\sim 10\%$ of the total dark matter relic abundance resides in the heavier component. This requirement follows from the preferred values of the particle lifetime. The fraction of particles that have decayed is given by
\begin{equation}
\frac{Y_1}{Y_\text{DM}} = 1 - \exp\left(-\frac{t_0}{\tau}\right),
\end{equation}
where $t_0\approx 4.3 \times 10^{17}$ s is the age of the Universe and $\tau$ is the lifetime of $N_1$. Using a typical value of $\tau=4 \times 10^{18}$ s, one finds that, to address the S$_8$ tension, roughly 10\% of the heavier dark matter component has to have decayed. If the decaying component constitutes only a fraction of the total dark matter, a shorter lifetime can compensate for the smaller fraction, provided the initial fractional abundance is large enough.
The results of an exemplary computation are shown in Fig.~\ref{fig:conversion} for a mass ratio $m_\phi / m_1~=~10$. Here we select two representative values of $y_{11}$, that are motivated by achieving the correct DM relic abundance, and study which fraction of the initial abundance survives the conversion process as a function of the ratio $y_{22}/y_{11}$. As can be seen, the fraction of particles that do not undergo conversion via annihilation depends sensitively on the Yukawa couplings. This is expected since the cross section scales as $\sigma \propto y_{11}^2 y_{22}^2$.  The results show that conversion is quite efficient for a relatively large $y_{11}\sim10^{-3}$; in this case, a modest hierarchy of $y_{22}/y_{11}\lesssim 1/4$ is required to ensure that more the $10\%$ of the heavier sterile neutrino $N_1$ survives until the late Universe. In order to retain the majority of $N_1$ particles $y_{22}/y_{11}\lesssim 1/10$ is needed in this case. For smaller $y_{11}$ conversion processes are less efficient and we find that $Y_1/Y> 0.1$ for all $y_{22}<y_{11}$ while $y_{22}/y_{11}\lesssim 1/2$ is sufficient to ensure that more than half of the $N_1$ population remains. In order to ensure that $N_1$ constitutes the bulk of the dark matter in the Universe we will therefore adopt a mildly hierarchical choice for the Yukawas and concentrate on $y_{22}/y_{11}=\mathcal{O}(0.1)$.

\subsection{Decay}
\label{sec:width}
\begin{figure}[t!]
\centering
\begin{tikzpicture}
\begin{feynman}
\vertex (a) {\(N_1\)};
\vertex[right=2cm of a] (b);
\vertex[above right=1.5cm of b] (f1) {\(N_2\)};
\vertex[below right=1.5cm of b] (z);
\vertex[above right=1.5cm of z] (f2) {\(N_2\)};
\vertex[below right=1.5cm of z] (f3) {\(N_2\)};
\diagram* {
  (a) -- [fermion] (b) -- [fermion] (f1),
  (b) -- [scalar, edge label=\(\phi\)] (z),
  (z) -- [fermion] (f2),
  (z) -- [anti fermion] (f3),
};

\end{feynman}
\end{tikzpicture}
\caption{Representative Feynman diagram for the decay of the heavier sterile $N_1$ into three $N_2$.}
\label{fig: decay of sterile neutrino into active neutrinos}
\end{figure}
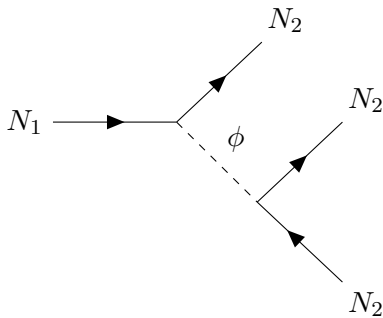
The heavier sterile neutrino $N_1$ can decay to three lighter sterile neutrinos $N_2$ if $m_1\geq 3 m_2$ and the off-diagonal Yukawa $y_{12}\neq0$\footnote{Alternatively, the decay to two lighter sterile neutrinos and an active neutrino is also possible. This process is suppressed by the mixing angle and constitutes a subleading contribution unless $y_{12}/y_{11}\lesssim \sin^2( \theta)$.}, see Fig.~\ref{fig: decay of sterile neutrino into active neutrinos} for a representative Feynman diagram. 
The width of this process is given by
\begin{align}
    \Gamma = \frac{1}{64 \pi^3 m_1} \int d E_2 \int d E_3 |M|^2,
\end{align}
where $E_2$ and $E_3$ are the energies of two of the final state particles. $|M|^2$ denotes the squared matrix element, which can be given in terms of $E_2$ and $E_3$ since we are averaging over the spin orientation.

Using the general condition for
the physical three-body phase space outlined in e.g. \cite{byckling1973particle} the limits of integration can be identified as
 $E_{3,min}=m_2$, $E_{3,max}=(m_1^2- 3 m_2^2)/(2 m_1)$ and
\begin{align}
E_{2,mim}=\frac{1}{2} \left(m_1 -E_3\pm\sqrt{\frac{(E_3-m_2) (E_3+m_2)
   \left(-2 E_3 m_1+m_1^2-3 m_2^2\right)}{m_1^2+m_2^2 -2 E_3
   m_1}}\right)\,.
\end{align}
While the first integral can be performed analytically, the second one does not have a solution in terms of elementary functions. 
This is a known issue that persists for all three-body phase space integrals with three massive particles. The problem can easily be solved numerically, but this turns out to be unnecessary in the regime we are interested in, i.e. $3 m_2 \approx m_1$, since a simple non-relativistic approximation works well here. 
Evaluating $|M|^2$ at the mean energy $\bar{E}_3=\bar{E}_2=m_1/3$ and using the non-relativistic approximation for the massive three-body phase space $R_{3\mbox{\tiny NR}}=\frac{\pi^3}{6 \sqrt{3}}(m_1-3m_ 2)^2$ \cite{byckling1973particle} we arrive at 
\begin{align}
\Gamma=\frac{y_{22}^2 y_{12}^2 }{864 \sqrt{3} \pi^2} \frac{m_1^5}{m_\phi^4} q^3 (2-q)^2,
\end{align}
where $q=(m_1-3m_2)/m_1$ is the fractional energy release of the decay.
We find that this approximation reproduces the numerical solution to better than $10\%$ for $q\lesssim 0.1$, which covers the full range of interest here.

In addition, we need to identify the mass splitting that leads to the typical velocity of the decay products required to suppress structure on the right scales. In order to capture the effect on structure formation, \cite{Fuss:2022zyt} proposed to map the three-body case to the two-body one by equating the pressure-like quantity 
\begin{align}
    \mathcal{P}=\left\langle \frac{p^2_{3}}{3 E_{3}} \right\rangle= \frac{1}{\Gamma} \int d E_{3} \frac{p^2_{3}}{3 E_{3}} \frac{d \Gamma}{d E_{3}},
\end{align}  which was used to approximate the effect of free-streaming on the cosmological perturbations in the modified CLASS code of \cite{FrancoAbellan:2021sxk}.
Plugging in the mean energy $\bar{E}_3=m_1/3$ leads to $\mathcal{P}_{3\mbox{\tiny body}} \approx \frac{2-q}{12} q\, m_1$, which reproduces the numerical results to better than $20\%$ for any value of $q$ and matches the numerical results for $q \rightarrow 0$. 
Comparing with the two-body result, $\mathcal{P}_{2\mbox{\tiny body}}  = \epsilon^2/(3-3\epsilon)\,m_1$ \cite{Fuss:2022zyt}, leads to the  preferred range $3.3 \times 10^{-4}\leq q \leq 3.7 \times 10^{-2}$. Note that this quantity is smaller than its equivalent in the two-body case since $\langle p_3^2/3 E_3 \rangle \propto q$ instead of $\propto \epsilon^2$. 
This can be understood by considering the different kinematics. In a two-body decay to a massive daughter particle and radiation, the bulk of the released kinetic energy is taken by the radiation since momentum conservation enforces equal momentum for the two final-state particles. In a decay to three identical massive particles, on the other hand, each particle has on average one third of the available kinetic energy.

\section{Parameter space for decaying sterile neutrinos}
\label{sec:our_decays}

 We are now in the position to combine the results for the dark matter relic density from Sec.~\ref{sec:DM_production} with the lifetime and the preferred range of $q$ from the previous subsection. We demand that the correct dark matter relic abundance is produced in the heavier states $N_1$ by the interplay of oscillations and self-interactions in the early Universe and that its lifetime and the energy release of the decay match the values required to achieve S$_8=0.77$, i.e. $\tau= 4 \-- 10 \times 10^{18}$ and $3.3 \times 10^{-4}\leq q \leq 3.7 \times 10^{-2}$.  
 Besides the relic density and S$_8$, there are additional observables that impose constraints on the parameter space.

 First, sterile neutrinos undergo a phenomenologically important one-loop decay $N_1\rightarrow \nu_a + \gamma$ due to their mixing with the active neutrinos. The corresponding decay rate is given by \cite{Pal:1981rm}
\begin{equation}
    \Gamma_{N_1 \rightarrow \nu_a \gamma}  \approx 1.35 \times 10^{-29}  \left(\frac{\sin^2(2 \theta)}{10^{-8}}\right) \left( \frac{m_1}{1 ~ \text{keV}}\right)^5 ~\text{s}^{-1}.
\end{equation}  
In this two-body decay, the energy of the resulting photon is $m_1/2$ which, for $\mathcal{O}(\text{keV})$ neutrinos, lies in the X-ray band. Such X-ray photons are searched for by current and future X-ray telescopes, providing some of the strongest observational constraints on sterile neutrino dark matter. 
Here, we use a compilation of the limits presented in \cite{Boyarsky:2007ge,Horiuchi:2013noa,Roach:2019ctw,Foster:2021ngm} which bound the mixing angle from above in a way that is independent of the details of the new interaction. These limits are shown in gray in the figures. 

Second, since sterile neutrinos are produced while they are still relativistic, they can have an impact on structure formation that resembles that of warm dark matter (WDM). 
The free streaming of the decoupled neutrinos can lead to a suppression in the matter power spectrum on small scales that can be constrained with current observational data\footnote{The relevant scales here are more than one order of magnitude smaller than those that contribute significantly to S$_8$ and a suppression of the power spectrum on these scales has a negligible impact on it.}, e.g. \cite{Dekker:2021scf,Hsueh:2019ynk, Gilman:2019nap}. 
In general, a precise evaluation of structure formation bounds requires detailed cosmological simulations. 
Here, however, we adopt the simplified approach from \cite{Astros:2023xhe} and refer to it for a full discussion. In this approach, we match the results for a thermal WDM candidate to our model. 
Typically, structure formation bounds are reported in terms of the minimal mass that the warm relic can have. This can be translated to a bound in the maximum root-mean-squared velocity that DM particles have today \cite{Bode:2000gq, Barkana:2001gr} 
\begin{equation}
    v_{\text{rms}} \approx 0.04 \left( \frac{\Omega h^2}{0.12}\right)^{1/3} \left( \frac{m_{\text{WDM}}}{1 ~\text{keV}}\right)^{-4/3} \text{km}~ \text{s}^{-1}.
\end{equation}
We take the conservative limit on the mass of the thermal relic of $m_{\text{WDM}} \gtrsim 1.9 \,\text{keV}$ derived in \cite{Garzilli:2019qki} from Lyman-$\alpha$ observations that corresponds to $v_\text{rms} \lesssim 16 \,\text{m/s}$. This bound is shown in light purple in the figures. Note that the bounds are stronger for the small mass ratio, since for light mediators the bulk of the neutrinos are produced at late times, so they inherit the heating of the SM bath due to the decrease of the number of relativistic degrees of freedom. In both cases, the bounds relax for smaller mixing angles due to the presence of partial thermalization.   

Finally, self-interactions can also influence structure formation, as DM particles remain collisional even after chemical decoupling. Such self-interactions lead to a suppression of the matter power spectrum due to pressure support \cite{deLaix:1995vi,Atrio-Barandela:1996suw,Hannestad:2000gt}. The resulting effect is analogous to free-streaming, in that it inhibits the formation of structures below a characteristic scale. The amount of suppression is ultimately controlled by the time of kinetic decoupling which is set by the self-scattering cross section \cite{Egana-Ugrinovic:2021gnu}. 
For the mediator masses considered here, the impact of self-interactions is modest, see e.g. \cite{Astros:2023xhe}.

\begin{figure}[t]
\includegraphics[width = \linewidth]{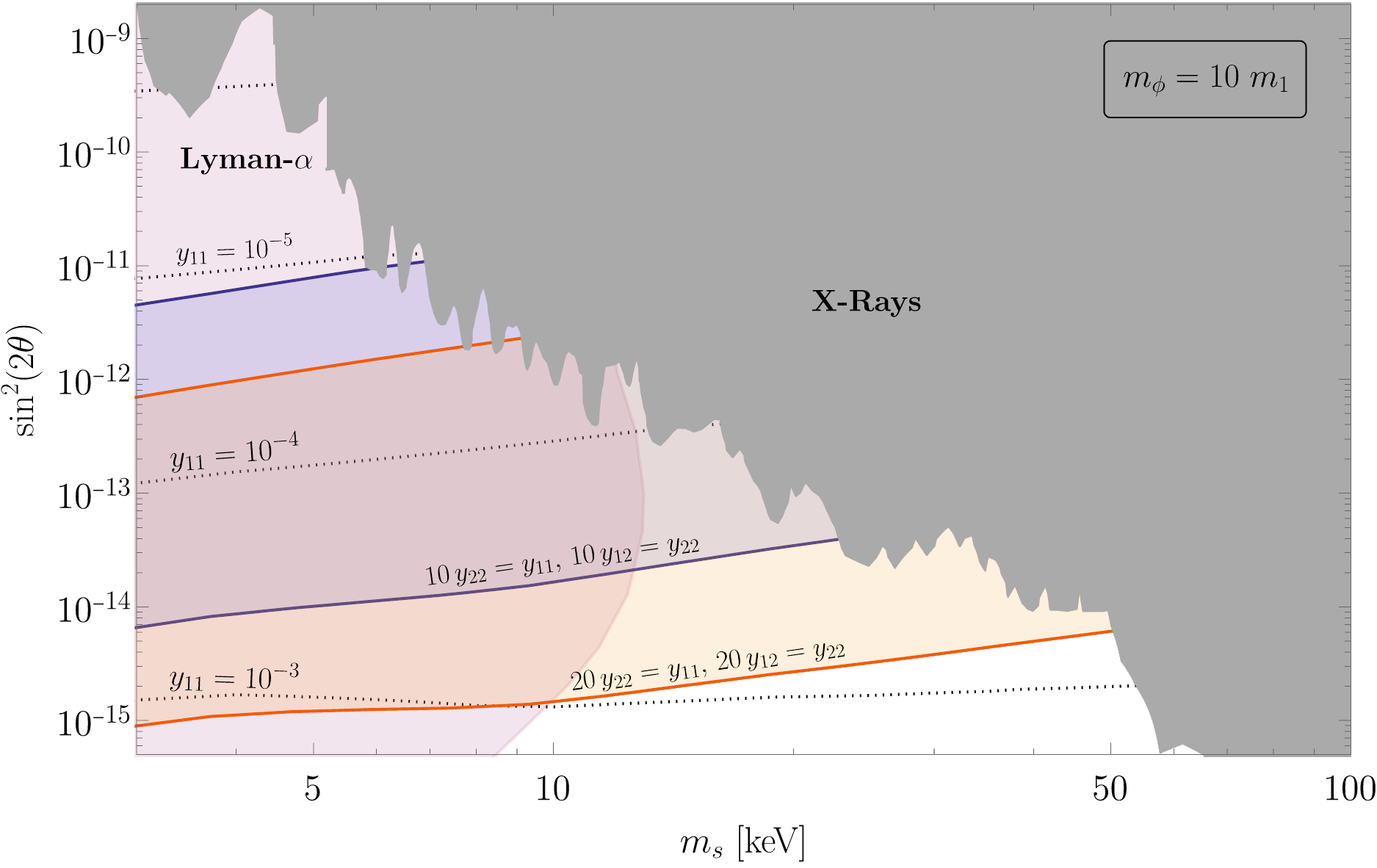} 
\caption{Parameter space in the $m_1 - \sin^2(2\theta)$ plane for a fixed mediator mass ratio $m_\phi = 10 m_1$. The black dotted contours indicate the values of $y_{11}$ for which, in the absence of the other Yukawa couplings, the relic abundance of $N_1$ matches the observed value. The allowed parameter space is limited by X-ray constraints (gray) and Lyman-$\alpha$ bounds (light purple), see text for details. Between the solid color lines (blue and yellow) the lifetime of $N_1$ is of the order of $10^{18}\,\text{s}$ and $3.3 \times 10^{-4}\leq q\leq 3.7 \times 10^{-2}$ as preferred to solve the S$_8$ tension. 
}
\label{fig: light parameter space}
\end{figure}

\begin{figure}
    \centering
    \includegraphics[width=\linewidth]{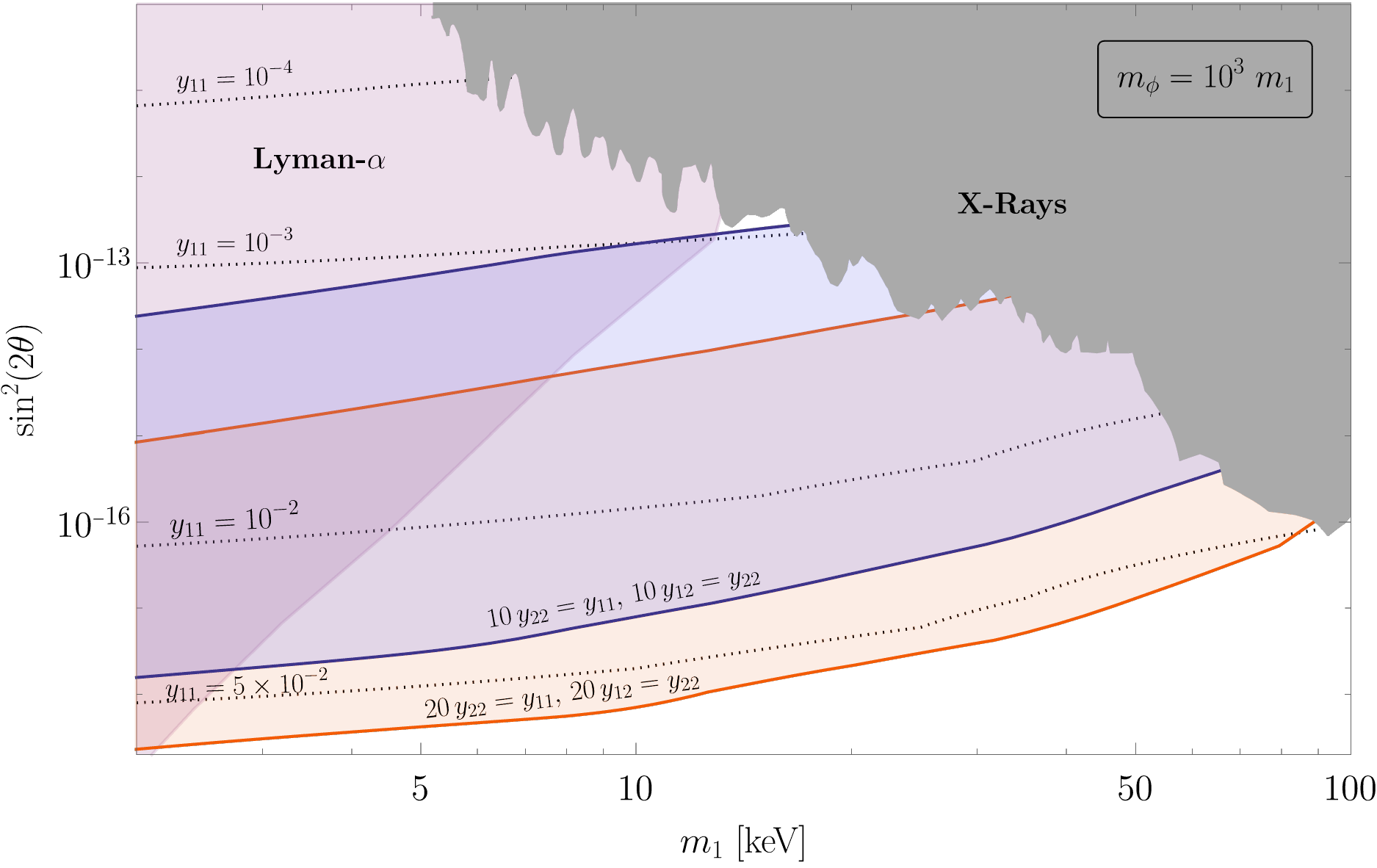}
    \caption{Parameter space available in the $\sin^2(2\theta) - m_1$ plane for a fixed mediator mass $m_\phi = 10^3\, m_1$. The black dotted contours indicate the values of $y_{11}$ for which, in the absence of the other Yukawa couplings, the relic abundance of $N_1$ matches the observed value. We also show a compilation of X-ray constraints (gray) and Lyman-$\alpha$ bounds (light purple). Between the solid color lines (blue and yellow) the lifetime of $N_1$ is of the order of $10^{18}\,\text{s}$ and $3.3 \times 10^{-4}\leq q\leq 3.7 \times 10^{-2}$ as preferred to solve the S$_8$ tension. 
    }
    \label{fig: heavy parameter space}
\end{figure}

The results are presented in Fig.~\ref{fig: light parameter space} and Fig.~\ref{fig: heavy parameter space} where we show the available parameter space of the model in the $m_1 - \sin^2(2\theta)$ plane for two exemplary mass ratios of $m_\phi = 10 \,m_1$ and $m_\phi = 10^3 \, m_1$.
For  each point in the $m_1 - \sin^2(2\theta)$ plane the Yukawa coupling $y_{11}$ was chosen in such a way that the produced $N_1$ particles comprise all of the DM after being produced by the mechanism described in Sec.\,\ref{sec: boosted production}. These values of $y_{11}$ are shown as dotted black lines in the figures. 
Additionally, in the region between the solid blue and yellow lines the lifetime of the $N_1$ particles is $\tau = 5 \times 10^{18} \,\text{s}$ and $3.3 \times 10^{-4}\leq q \leq 3.7 \times 10^{-2}$ as preferred by the S$_8$ tension. The blue region corresponds to a mild hierarchy of the Yukawas of $y_{22} = y_{11}/10$ and $y_{12} = y_{22}/10$, while the yellow region corresponds to the choice $y_{22} = y_{11}/20$ and $y_{12}= y_{22}/20.$ As we go down within each region the Yukawa coupling increases by approximately one order of magnitude. In order to keep the lifetime in the preferred window we therefore need to decrease the value of q accordingly. 
Thus, the upper edge of the shaded region is characterized by $q=3.7 \times 10^{-2}$ while the lower one has $q=3.3 \times 10^{-4}$.
As can be seen by comparing Fig.~\ref{fig: light parameter space} and Fig.~\ref{fig: heavy parameter space}, decays with the properties that lead to a modest reduction of S$_8$ can be achieved for a wide range of $m_\phi$ masses.

Outside these regions, the model accounts for the entirety of DM, but the combined requirements on the lifetime and on $q$ that lead to S$_8=0.77$ cannot be fulfilled simultaneously.  It is instructive to consider the effect of giving up one of these restrictions. Let us consider fixing the lifetime first. Going above the upper line then requires a larger value of $q$. This shifts the suppression of the power spectrum to larger scales. Since observations agree well with expectations in this regime, there are strong constraints here and such a solution to S$_8$ is not allowed. Going below the lower line, on the other hand, can be achieved by lowering $q$. This shifts the suppression to smaller scales that are less well tested and for sufficiently low $q$, the model becomes indistinguishable from stable $N_1$ dark matter.
In contrast, for a fixed value of $q$, going above the upper line corresponds to an increase in the lifetime. This reduces the effect on structure formation, and we again recover the stable limit if we move too far from the line. Going below the lower line, however, is accompanied by a shorter lifetime and more $N_1$ particles will have decayed. This leads to a stronger suppression of the power spectrum on roughly the same scale, and S$_8$ drops further. This is in conflict with observations and, therefore, this direction is strongly constrained. Taken together, this implies that the model is not simply allowed or disfavored outside the band. Instead, the results depend on $q$ and both allowed and disfavored regions can be found on both sides.

\section{Summary and Conclusions}
\label{sec:summary}

There are intriguing hints that the amplitude of the matter power spectrum on scales of 8~Mpc might be reduced compared to the expectation from $\Lambda$CDM. One possibility to realize such a deviation from concordance cosmology is to give up the cold dark matter paradigm and consider a non-trivial late-time dynamics in the dark sector. A solution, that has been identified in the literature, is a decaying dark matter model where at least one of the resulting daughter particles is produced with a non-relativistic velocity. Most cosmological studies only consider a simple effective parametrization of these decays in terms of the lifetime and the fractional momentum of the daughter particle. 

It is interesting to study how such a scenario can be realized in a full dark matter model. This allows us to track the evolution of the dominant dark matter population back to the early Universe and to establish links to other observables. Here, we have investigated a simple model based on two interacting sterile neutrinos. While only one of them mixes with the active neutrinos, both interact with a new scalar singlet with a general Yukawa matrix. 
The new interactions enable a boosted production of $N_1$ in the early Universe and allow them to decay into three lighter steriles $N_2$ at late times. The particles produced by the decay take the role of the non-relativistic daughter in the two-body decay model and can reduce S$_8$. 
To the best of our knowledge this is the first study of an all massive three-body decay in this context.

To assess whether this mechanism can quantitatively reproduce the desired suppression of structure formation, we mapped the results of a cosmological simulation for the two-body decay to the massive three-body decay using the pressure-like quantity $\mathcal{P}=\left\langle p^2_{3}/3 E_{3}\right \rangle  $ and found that $(m_1-3 m_2)/m_1\sim 10^{-3} $ leads to a reduction of the amplitude of the matter power spectrum on the right scales.  Imposing this restriction on the masses leads to a significant suppression of the phase space and increases the stability of the heavier sterile neutrino. Nevertheless, we find that lifetimes of the order $10^{18}\-- 10^{19}$ s, as required by a reduction of $S_8$, can be achieved quite naturally for a large range of $m_\phi/m_1$ ratios if we assume a mild hierarchy between a dominant Yukawa coupling $y_{11}$ and the other couplings $y_{22}$ and $y_{12}$. This solution can then be confronted with other constraints, such as limits from X-ray searches and observations of structure on smaller scales as tested by e.g. the Lyman-$\alpha$ forest. Together these exclude both very light and heavy sterile neutrinos and leave a window from $\sim10\--100$ keV where our sterile neutrino can fulfill all constraints simultaneously.

\acknowledgments
MD and SV acknowledge support by the DFG via the individual research grant Nr.
496940663. SV thanks the Mainz Institute of Theoretical Physics (MITP) for hospitality during the completion of this work. LG acknowledges
support from the Dutch Research Council (NWO) under
project number VI.Veni.222.318 and from Charles University through project number PRIMUS/24/SCI/013.

\bibliographystyle{JHEP}
\bibliography{biblio.bib}




\end{document}